\documentclass[prd,superscriptaddress,showpacs,nofootinbib,amsmath,amssymb,aps,10pt]{revtex4}

\usepackage{bm}
\usepackage{amsfonts}
\usepackage{latexsym}
\usepackage[latin1]{inputenc}
\usepackage{graphicx}
\usepackage{amsmath}
\usepackage{palatino}
\usepackage{mathpazo}
\usepackage{epstopdf}
\usepackage{booktabs}
\usepackage{dcolumn}
\usepackage{array}

\usepackage{wrapfig}
\usepackage{wasysym}

\def\jnl@style{\it}
\def\aaref@jnl#1{{\jnl@style#1}}

\def\aaref@jnl#1{{\jnl@style#1}}

\def\aj{\aaref@jnl{AJ}}                   
\def\apj{\aaref@jnl{ApJ}}                 
\def\apjl{\aaref@jnl{ApJ}}                
\def\apjs{\aaref@jnl{ApJS}}               
\def\apss{\aaref@jnl{Ap\&SS}}             
\def\aap{\aaref@jnl{A\&A}}                
\def\aapr{\aaref@jnl{A\&A~Rev.}}          
\def\aaps{\aaref@jnl{A\&AS}}              
\def\mnras{\aaref@jnl{Mon.~Not.~Roy.~Astron.~Soc.}}             
\def\prd{\aaref@jnl{Phys.~Rev.~D}}        
\def\prc{\aaref@jnl{Phys.~Rev.~C}}  
\def\prl{\aaref@jnl{Phys.~Rev.~Lett.}}    
\def\qjras{\aaref@jnl{QJRAS}}             
\def\skytel{\aaref@jnl{S\&T}}             
\def\ssr{\aaref@jnl{Space~Sci.~Rev.}}     
\def\zap{\aaref@jnl{ZAp}}                 
\def\nat{\aaref@jnl{Nature}}              
\def\aplett{\aaref@jnl{Astrophys.~Lett.}} 
\def\apspr{\aaref@jnl{Astrophys.~Space~Phys.~Res.}} 
\def\physrep{\aaref@jnl{Phys.~Rep.}}      
\def\physscr{\aaref@jnl{Phys.~Scr}}       
\def\commat{\aaref@jnl{Comm.~Math.~Phys.}}              
\def\science{\aaref@jnl{Science}}               
\def\cqg{\aaref@jnl{Classical Quant.~Grav.}}            
\def\jpcs{\aaref@jnl{JPCS}}                                     
\def\ijmpd{\aaref@jnl{Int.~J.~Mod.~Phys.~D}}                    
\def\grg{\aaref@jnl{Gen.~Relat.~Gravit.}}               
\def\rpp{\aaref@jnl{Rep.~Prog.~Phys.}}          
\def\npa{\aaref@jnl{Nucl.~Phys.~A}}        
\def\lrr{\aaref@jnl{Living Rev.~Rel.}}                   
\def\jcap{\aaref@jnl{J.~Cosmology Astropart.~Phys.}}    
\def\rmp{\aaref@jnl{Rev.~Mod.~Phys.}}   

\allowdisplaybreaks[1]

\begin{document}

\title{Moment of inertia of neutron star crust in alternative and modified theories of gravity}

\author{Kalin V. Staykov}
\email{kstaykov@phys.uni-sofia.bg}
\affiliation{Department of Theoretical Physics, Faculty of Physics, Sofia University, Sofia 1164, Bulgaria}

\author{K. Yavuz Ek\c{s}i}
\email{eksi@itu.edu.tr}
\affiliation{ Faculty of Science and Letters, Department of Physics, 34469 Maslak, Istanbul, Turkey }

\author{Stoytcho S. Yazadjiev}
\affiliation{Department of Theoretical Physics, Faculty of Physics, Sofia University, Sofia 1164, Bulgaria}

\author{M. Metehan T{\"u}rko{\u g}lu}
\affiliation{ Faculty of Science and Letters, Department of Physics, 34469 Maslak, Istanbul, Turkey }

\author{A. Sava{\c s} Arapo{\u g}lu}
\affiliation{ Faculty of Science and Letters, Department of Physics, 34469 Maslak, Istanbul, Turkey }


\begin{abstract}

The glitch activity of young pulsars arises from the exchange of angular momentum between the crust and the interior of the star.
Recently, it was inferred that the moment of inertia of the crust of a neutron star is not sufficient to explain the observed glitches.
Such estimates are presumed in the Einstein's general relativity in describing the hydrostatic equilibrium of neutron stars.
The crust of the neutron star has a space-time curvature of 14 orders of
magnitude larger than that probed in solar system tests.
This makes gravity the weakest constrained physics input in the crust related processes.
We calculate the ratio of the crustal to the total moment of inertia of neutron stars in the scalar-tensor theory of gravity and the non-perturbative $f({\cal R})={\cal R}+ a {\cal R}^2$  gravity. We find for the former that  the crust to core ratio of the moment of inertia
does not change significantly from what is inferred in general relativity. For the latter we find that the
ratio increases significantly from what is inferred in general relativity in the case of high mass objects.
Our results suggest that the glitch activity of pulsars may be used to probe gravity models,  although the gravity models explored in this work are not appropriate candidates.

\end{abstract}
\maketitle
\date{}
\section{Introduction}

Rotationally powered pulsars are strongly magnetized rapidly rotating neutron stars
spinning-down under the magnetic dipole radiation
torques  \cite{Pacini1967,Gold1968,Gunn1969,Ostriker1969,Pacini1968}. More than 2000 pulsar have been discovered to date \cite[see][for a review]{Lyne2012}.
Some of these objects, predominantly the younger ones, show occasional and abrupt spin-up events called glitches \cite{Radhakrishnan1969,Reichley1969}
of the size $\Delta \Omega/\Omega \sim 10^{-9}-10^{-6}$ where $\Omega$ is the rotational angular velocity of the neutron star \cite[see][for observational data]{Espinoza2011,Yu2013}. Of the more than 2000 pulsars about 160 were observed to glitch with a total of $\sim 400$ glitches. The glitches are generally accompanied by a sudden change of the spin-down rate $|\Delta \dot{\Omega}/\dot{\Omega} | \sim 10^{-4}-10^{-2}$. These glitches tend to relax back to their original pre-glitch state values.
The long time-scale (days to years) of the post-glitch relaxation implies the presence of a superfluid interior \cite{Baym1969} \cite[see][for a review]{Page2013}.

The small glitches, as observed in the Crab pulsar, can be explained with the starquake model \cite{Ruderman1969} in which the solid crust resisting the reduction of the oblateness of the figure of the spun-down star breaks at a critical strain.   The starquake model predicts a time interval between the large glitches much longer than observed e.g.\ in the Vela pulsar \cite{Baym1971}. More importantly the energy dissipation in such a large glitch would lead to an observable change in the surface temperature exceeding the bounds \cite{AtakanGurkan2000}. Hence the starquake model can not be the mechanism underlying the Vela-like large glitches which comprise the majority of the observed glitches.

Such large glitches are explained by the exchange of angular momentum between the superfluid component permeating the inner crust of a neutron star and the relatively slower rotating outer crust.
The spin-down torques due to the magnetic dipole radiation act on the outer crust and slow it down.
The neutron superfluid permeating the inner crust follows the spin-down of the outer crust as they are weakly coupled by mutual friction forces \cite{Andersson2006}.
The angular momentum of the superfluid is carried by an array of quantized vortices \cite{Packard1972,Anderson1975} and its slow-down is achieved by the radial motion of the vortices away from the rotation axis \cite{Alpar1984}. The pinning of the vortices to the nuclei precludes their radial motion preventing the spin-down of the neutron superfluid, and results in a lag in which the superfluid rotates more rapidly than the crust \cite{Anderson1975,Alpar1984}.
This lag induces a Magnus force acting on the vortices and produces a stress on the crust.
When the lag reaches a critical threshold the vortices are suddenly unpinned in an avalanche transferring angular momentum from the rapidly rotating neutron superfluid to the crust causing its observed spin-up \cite{Anderson1975,Melatos2008}. The post-glitch relaxation may result from repinning	 of the unpinned vortices to new pinning sites and restart to ``creep'' in steady state \cite{Alpar1984}.

Superfluid neutrons of the inner crust, being in Bloch states of the crust lattice, have an effective mass
larger than the bare neutron mass \cite{Chamel2005,Chamel2012}. This so called `entrainment' results with a less mobile crustal superfluid because of the reduced ratio of the effectively free
neutron superfluid that can store and exchange angular momentum with the lattice \cite{Chamel2006,Andersson2012,Chamel2013}.
In order to address this problem authors of ref.\ \cite{Guegercinouglu2014} invoked the contribution of the toroidal magnetic flux region in the outer core of the star to the moment of inertia of the crust.
 Piekarewicz et al.\ \cite{Piekarewicz2014}, on the other hand, considered nuclear physics processes that could lead to a thicker crust.

The arguments leading to the problem of the deficiency of the crustal moment of inertia for the glitch phenomena and explanations for addressing the problem rely on the calculations of the structure of neutron stars within the framework of Einstein's general relativity (GR). Although GR is tested to a precision of $10^{-5}$ in the solar system tests and found to be in excellent agreement with measurements \cite[see][for a review]{Will2014}, the presumption of the validity of GR for describing the hydrostatic equilibrium of neutron stars is a remarkable extrapolation, 5 orders of magnitude in compactness ($\sim GM/c^2 R$) and 14 orders of magnitude in curvature ($\sim GM/c^2 R^3$) \cite{Psaltis2008}.
The equation of state (EoS) of the outer crust \cite{Baym1971a} and the inner crust \cite{Negele73} are fairly well understood \cite[see][for a review]{Chamel2008} where the gravity of the neutron star attains its strongest value \cite{Ekcsi2014}.
Gravity is then the least constrained physics input for the glitch activity in the crust and, thus, such phenomena may be used to constrain gravity models.
The purpose of this paper is to draw attention to the role of gravity in the crust by calculating, for the first time, the fraction of the crustal moment of inertia in different models of gravity.

\section{Method}

n this section we briefly  present the basic equations for $f({\cal R})$ and scalar-tensor theory (STT) of gravity, which are presented more thoroughly in \cite{Yazadjiev2014,Staykov2014}.

The mathematical equivalence between the $f({\cal R})$ theories and  STT  is exploited in this work. For mathematical simplicity it is useful to study $f({\cal R})$ and scalar-tensor theories in the so-called Einstein frame, however all final results we present here are transformed to the Jordan frame (the physical frame). The explicit form of the function $f({\cal R})$ we adopt  is $f({\cal R})={\cal R}+a{\cal R}^2$.

The $f({\cal R})$ theory action in Jordan frame is:

\begin{eqnarray} \label{eq:fR_actionJ}
S = \frac{1}{16 \pi G} \int d^4 x \sqrt{-g} \left({\cal R} + a{\cal R}^2\right) + S_{\rm matter} \left(g_{\mu\nu}, \chi\right),
\end{eqnarray}
where $S_{\rm matter}$ is the action of the matter fields collectively denoted by $\chi$. The general form of the STT action in Jordan frame is the following:

\begin{eqnarray}\label{eq:STT_actionJ}
S =  \frac{1}{16 \pi G} \int d^4 x \sqrt{-g} \left(F(\Phi){\cal R} - Z(\Phi)g^{\mu\nu}\partial_{\mu}\Phi\partial_{\nu}\Phi \right. \\ \left. - 2U(\Phi) \right) + S_{\rm matter} \left(g_{\mu\nu}, \chi\right), \nonumber
\end{eqnarray}
where $\Phi$ is the gravitational scalar field. $F(\Phi), Z(\Phi)$, and  $U(\Phi)$ are functions that govern the dynamics of the gravitational scalar. In order the gravitation to carry positive energy $F(\Phi) >0$, and the nonnegativity of the scalar field implies $2F(\Phi)Z(\Phi) + 3(dF(\Phi)/d\Phi)^2 \geq 0$. 
We are using the mathematical equivalence between  $f({\cal R})$ theories and  STT theories to rewrite the action (\ref{eq:fR_actionJ}) like an STT action. For mathematical simplicity we rewrite both actions in Einstein frame, using a conformal transformation $g_{\ast\mu\nu} = F(\Phi)g_{\mu\nu}$ and the new scalar field $\varphi$ defined by

\begin{equation}
\left(\frac{d\varphi}{d\Phi}\right)^2 = \frac{3}{4}\left(\frac{d\ln{F(\Phi)}}{d\Phi}\right)^2 + \frac{Z(\Phi)}{2F(\Phi)}.
\end{equation}
We also define

\begin{equation}
A(\varphi) = F^{-\frac{1}{2}}(\Phi), \; 2V(\varphi) = U(\Phi)F^{-2}(\Phi).
\end{equation}

Now we have one Einstein frame STT action describing  both theories:

\begin{eqnarray} \label{eq:action_E}
S = \frac{1}{16 \pi G^{\ast}} \int d^4 x \sqrt{-g^{\ast}} \left({\cal R}^{\ast} - g^{\ast  \mu\nu}\partial_{\mu}\varphi\partial_{\nu}\varphi \right. \\ \left.- V(\varphi) \right) + S_{\rm matter} \left(A^2(\varphi)g_{\ast  \mu\nu}, \chi\right). \nonumber
\end{eqnarray}

In (\ref{eq:action_E}) all quantities marked with asterisk are with respect to the Einstein frame metric $g^{\ast  \mu\nu }$.

We consider stationary and axisymmetric space-time as well as stationary and axisymmetric fluid and scalar field configuration.
The metric describing this space-time in Einstein frame has the explicit form
\begin{eqnarray}
& ds^2_{*}= - e^{2\phi(r)}dt^2 + e^{2\Lambda(r)}dr^2 + r^2(d\theta^2 +
\sin^2\theta d\vartheta^2 ) - \nonumber \\ & 2\omega(r,\theta)r^2 \sin^2\theta  d\vartheta dt,
\end{eqnarray}
where we kept only first-order terms in the angular velocity $\Omega$ (slow rotation approximation).

The  field equations in Einstein frame for both theories are as follows

\begin{eqnarray}
\frac{1}{r^2}\frac{d}{dr}\left[r(1- e^{-2\Lambda})\right]= 8\pi G
A^4(\varphi) \rho +   e^{-2\Lambda}\left(\frac{d\varphi}{dr}\right)^2
+ \frac{1}{2} V(\varphi), \label{eq:FieldEq1} \\
\frac{2}{r}e^{-2\Lambda} \frac{d\phi}{dr} - \frac{1}{r^2}(1-
e^{-2\Lambda})= 8\pi G A^4(\varphi) p +  e^{-2\Lambda}\left(\frac{d\varphi}{dr}\right)^2 - \frac{1}{2}
V(\varphi),\label{eq:FieldEq2}\\
\frac{d^2\varphi}{dr^2} + \left(\frac{d\phi}{dr} -
\frac{d\Lambda}{dr} + \frac{2}{r} \right)\frac{d\varphi}{dr}= 
 4\pi G
\alpha(\varphi)A^4(\varphi)(\rho-3p)e^{2\Lambda} + \frac{1}{4}
\frac{dV(\varphi)}{d\varphi} e^{2\Lambda}, \label{eq:FieldEq3}\\
\frac{dp}{dr}= - (\rho + p) \left(\frac{d\phi}{dr} +
\alpha(\varphi)\frac{d\varphi}{dr} \right), \label{eq:FieldEq4} \\
\frac{e^{\phi-\Lambda}}{r^4} \frac{d}{dr}\left[e^{-(\phi+ \Lambda)}r^4 \frac{d{\bar\omega}(r)}{dr} \right] = 
 16\pi G A^4(\varphi)(\rho + p){\bar\omega}(r) ,
\end{eqnarray}

where we have defined
\begin{equation}
\alpha(\varphi)= \frac{d\ln A(\varphi)}{d\varphi} \;\;\; {\rm and}\;\;\;  \bar\omega = \Omega - \omega.
\end{equation}
Here $p$ and $\rho$ are the pressure and energy density in the Einstein frame and they are connected to the Jordan frame ones via $\rho_{*}=A^{4}(\varphi)\rho$ and $p_{*}=A^{4}(\varphi) p$, respectively.

The above system of equations,  combined with the equation of state for the star matter and appropriate boundary conditions, describe the interior  and the exterior of
the neutron star. We obtain the exterior solutions by imposing the conditions $\rho=p=0$.

The boundary conditions that we employ are the natural ones, namely
at the center of the star we have $\rho(0)=\rho_{\rm c}, \Lambda(0)=0,$ $\frac{d{\bar\omega}}{dr}(0)= 0,$ and at infinity we have $\lim_{r\to \infty}\phi(r)=0, \lim_{r\to \infty}\varphi
(r)=0,$ $\lim_{r\to \infty}{\bar\omega}=\Omega$. The coordinate radius $r_{\rm S}$ of the star is determined by the condition  $p(r_{\rm S})=0$ and  the physical radius  is given by $R_{\rm S}= A[\varphi(r_{\rm S})] r_{\rm S}$.

Imposing a different form of the conformal factor $A(\varphi)$ and the potential $V(\varphi)$ we can examine $f({\cal R})$ theory and STT with the same set of equations.
In the case of ${\cal R}$-squared gravity they have the form
\begin{equation}
A^2(\varphi)=
e^{-\frac{2}{\sqrt{3}}\varphi} \;\;\;\alpha = -\frac{1}{\sqrt{3}}\;\;\; V(\varphi)= \frac{1}{4a}
\left(1-e^{-\frac{2\varphi}{\sqrt{3}}}\right)^ 2,
\end{equation}
and for STT with no potential we adopt
\begin{equation}
A(\varphi)=e^{\frac{1}{2}\beta\varphi^2} \;\;\;\alpha = \beta\varphi\;\;\; V(\varphi)= 0.
\end{equation}

The moment of inertia $I$ of a	 neutron star is defined in the standard way
\begin{equation}
I=\frac{J}{\Omega},
\end{equation}
where $J$ is the angular momentum.
For convenience in the numerical calculations, the integral expression
\begin{equation}
I= \frac{8\pi G}{3} \int_{0}^{r_S}A^4(\varphi)(\rho + p)e^{\Lambda - \Phi} r^4
\left(\frac{\bar\omega}{\Omega}\right)\, dr
\label{eq:I_integral}
\end{equation}
will be used.

From now on we shall use the dimensionless parameter $a\to a/R^2_{0}$  and the dimensionless moment of inertia
$I\to I/M_{\odot}R^2_{0} $ where $M_{\odot}$ is the solar mass and $R_{0}$ is one half the solar gravitational radius   $R_{0}=1.47664 \,{\rm km}$.

\section{Numerical results}

\begin{figure*}[]
\centering
\includegraphics[width=0.42\textwidth]{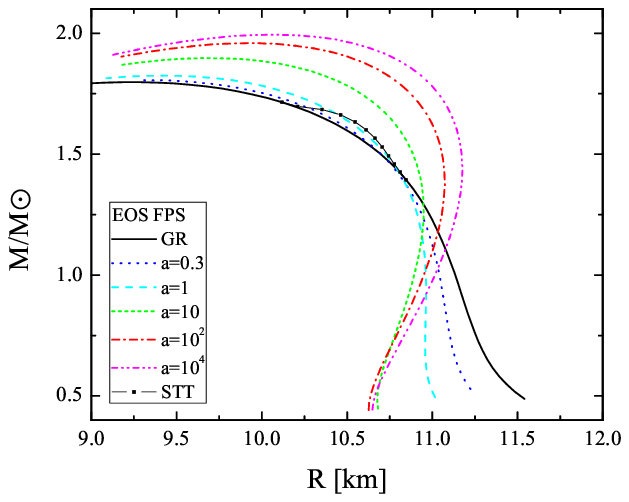}
\includegraphics[width=0.42\textwidth]{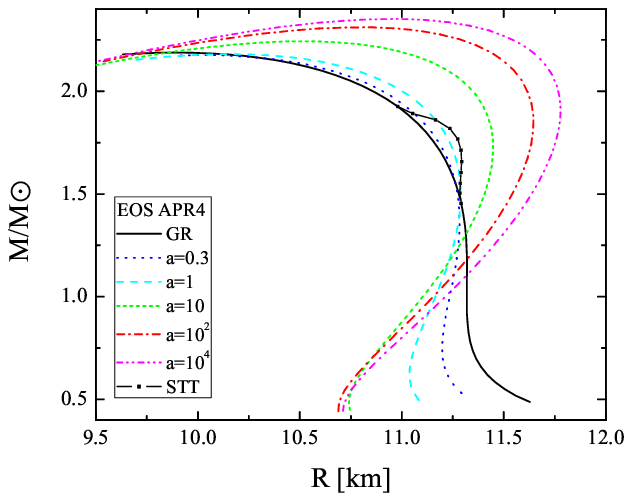}
\includegraphics[width=0.42\textwidth]{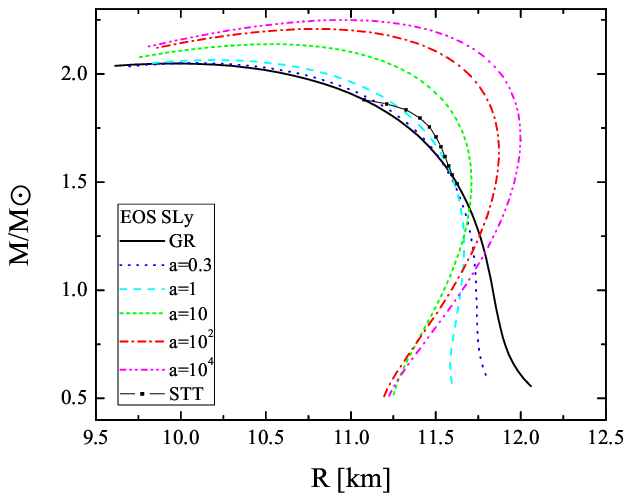}
\includegraphics[width=0.42\textwidth]{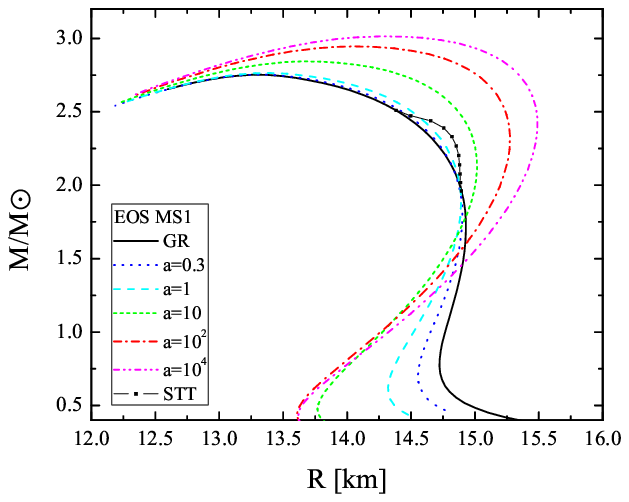}
\caption{Mass of radius relation for all examined EoS'. The case of GR is represented by a
solid black line, and the STT with $\beta=-4.5$ is given by black line with symbols, and the $\mathcal{R}^2$ gravity results are represented by lines with different patterns and different colours. }
\label{Fig_MR_all}
\end{figure*}

We solved, numerically, the hydrostatic equilibrium equations to investigate the behaviour of the ratio of $I_{\rm crust}$ to the total moment of inertia $I$ of neutron stars for four realistic
 EoS', namely FPS, SLy, APR4 and MS1. The first one, FPS, is a soft EoS with maximal mass less than the observed maximum mass of $2~M_{\odot}$ \cite{Demorest10,Antoniadis2013}.
SLy and APR4 have maximal masses slightly above $2~M_{\odot}$, and the last one, MS1, is the stiffest EoS---it gives highest masses and biggest radii. Thus a wide range of stiffness is covered by these choices of EoS'. In Fig. \ref{Fig_MR_all} we present the mass of radius relation for all mentioned EOS. The results are presented for GR, and for both alternative theories we are investigating. We define the crust of the neutron star as the layer with density lower than $1.5 \times 10^{14}~{\rm g~cm^{-3}}$ \cite{FPS}. We examined different values for this critical density and saw that the precise value does not affect our results and conclusions.

 \subsection{Scalar-tensor theory of gravity with vanishing potential}

We first investigate, numerically,  the ratio $I_{\rm crust}/I$ for neutron stars moment of inertia in scalar-tensor theories of gravity with vanishing potential \cite{Damour1993}.
The relativistic stellar models in
scalar-tensor gravity can significantly deviate from the predictions of general relativity \cite{DeDeo2003}. Spontaneous scalarization is induced	
 for $\beta \lesssim -4.35$ \cite{Harada1998}. Yet observational studies on pulsar timing provides the constraint  $\beta \geqq -4.5$ \cite{Will2014,Antoniadis2013} .
The value of the parameter we adopt is $\beta=-4.5$, which is consistent with the observational results. We did the calculations for all EoS' and compared the results with the results within GR.

\begin{figure*}[]
\centering
\includegraphics[width=0.42\textwidth]{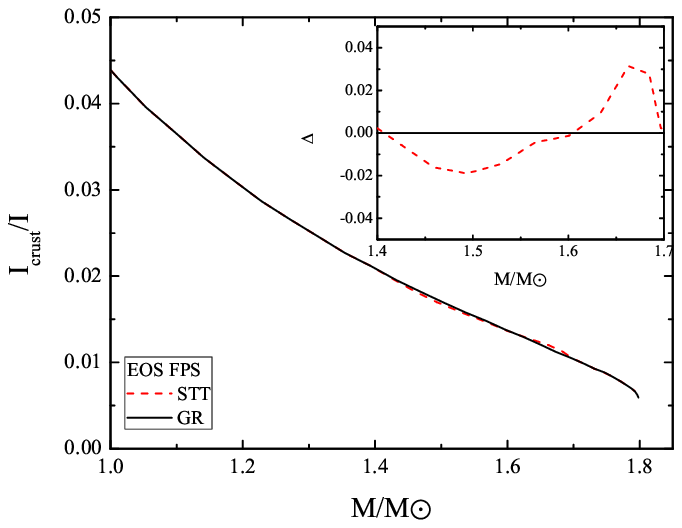}
\includegraphics[width=0.42\textwidth]{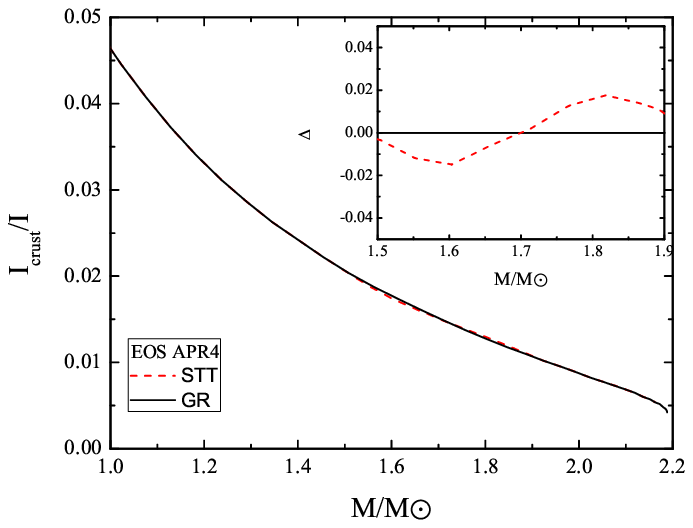}
\includegraphics[width=0.42\textwidth]{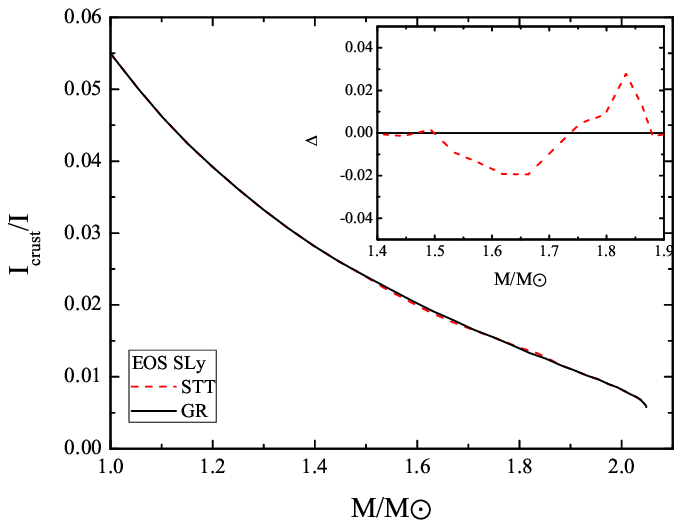}
\includegraphics[width=0.42\textwidth]{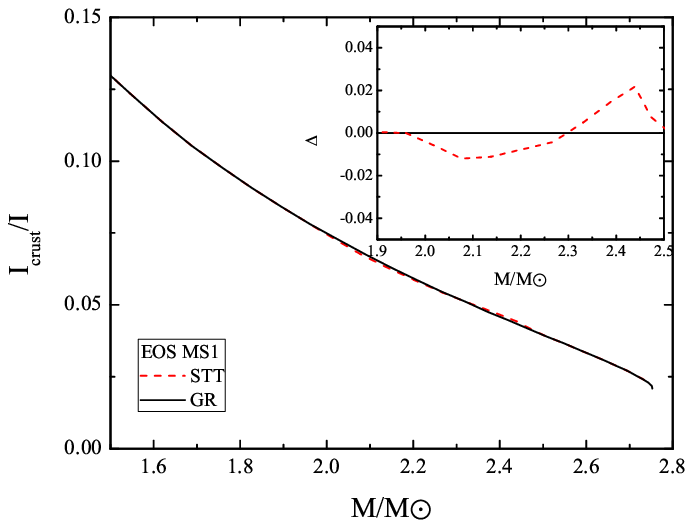}
\caption{Crust to the total moment of inertia ratio as a function of  mass for all EoS'. The case of GR is represented by a
solid black line, and the STT with $\beta=-4.5$ -- by a dashed (red) line. The inset shows the discrepancy between pure GR and the results in STT.}
\label{Fig_STT_all}
\end{figure*}

In Fig.  \ref{Fig_STT_all} we present the  ratio $I_{\rm crust}/I$  as
a function of the mass of the neutron star. The GR case is represented by a continuous black line and the STT one is in (red) dashed line. The chosen interval of masses is consistent with the  observations, but for representing the results for EoS MS1 we present a bigger mass range because the STT deviations from GR occur only for masses close to the maximal ones for this EoS.
For all EoS'	 we obtain qualitatively the same result: the $I_{\rm crust}/I$ ratio decreases rapidly with the mass of the neutron star.
For models with $M=0.5M_{\odot}$ the ratio is between $0.1\ \rm and\ 0.15$ (not presented in the graphs), and for models with maximal mass it is under $0.03$ for all EoS'. The insets shows the discrepancy  between GR and STT solutions. The discrepancy we define as
\begin{equation}
\Delta = \frac{\frac{I_{\rm crust}}{I}( STT) - \frac{I_{\rm crust}}{I} ( GR)}{\frac{I_{\rm crust}}{I} (GR)},
\end{equation}
The difference is much smaller than $3\%$  which is expected because the deviations from pure GR for $\beta = -4.5$ for non-rotating (slowly rotating) neutron stars are not large, for all EoS'. For the case of rapidly rotating neutrons stars, however, deviations in the moment of inertia increase \cite{Doneva2013}, therefore the differences for $I_{\rm crust}/I$ also increase as expected. As a result the deficiency in the crustal moment of inertia cannot be explained by the concrete STT examined in this paper.

\subsection{${\cal R}$-squared gravity}

We continue by investigating numerically  the behaviour of the crust to the total moment of inertia ratio  in ${\cal R}^2$ gravity. The observational constraints for the parameter $a$ are the following -  $a \lesssim 5 \ 10^{11} m^2$ \cite{Naef2010} or in dimensionless units $a \lesssim  10^{5}$.

\begin{figure*}[]
\centering
\includegraphics[width=0.42\textwidth]{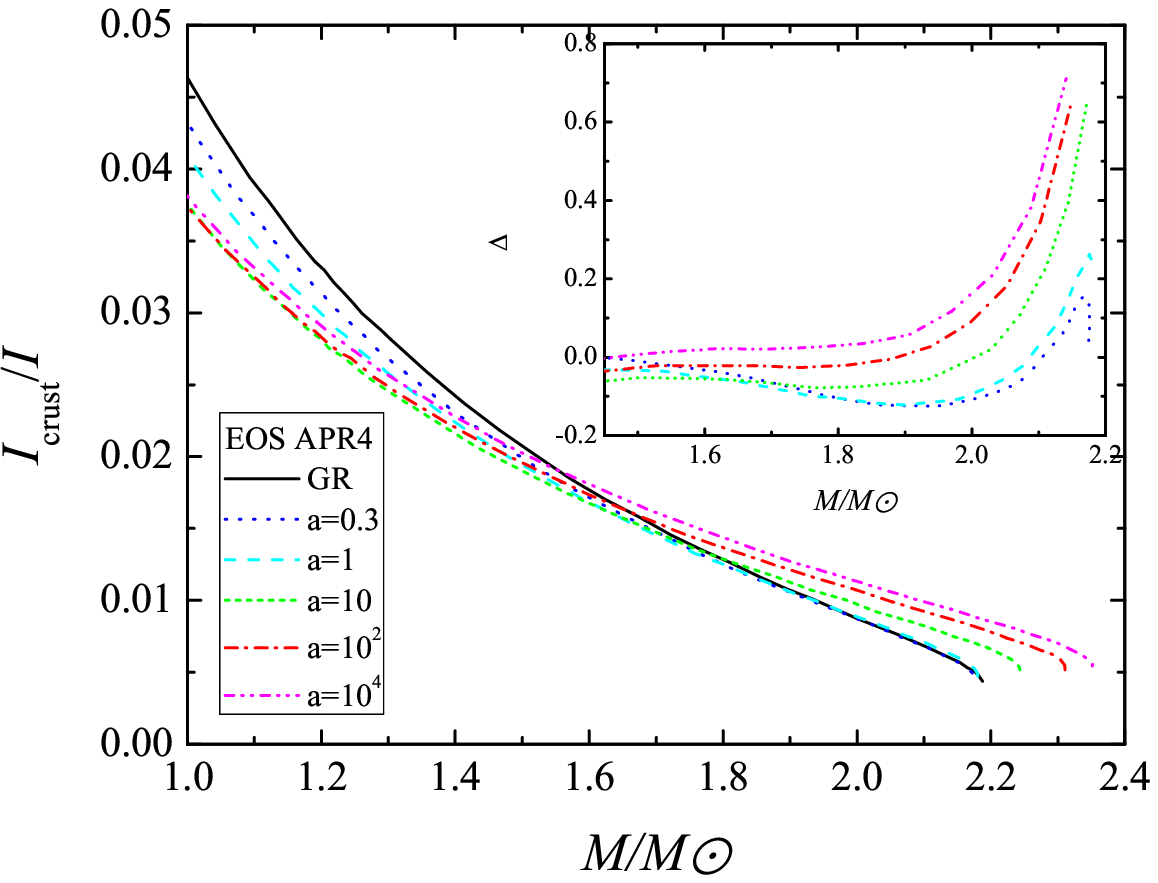}
\includegraphics[width=0.42\textwidth]{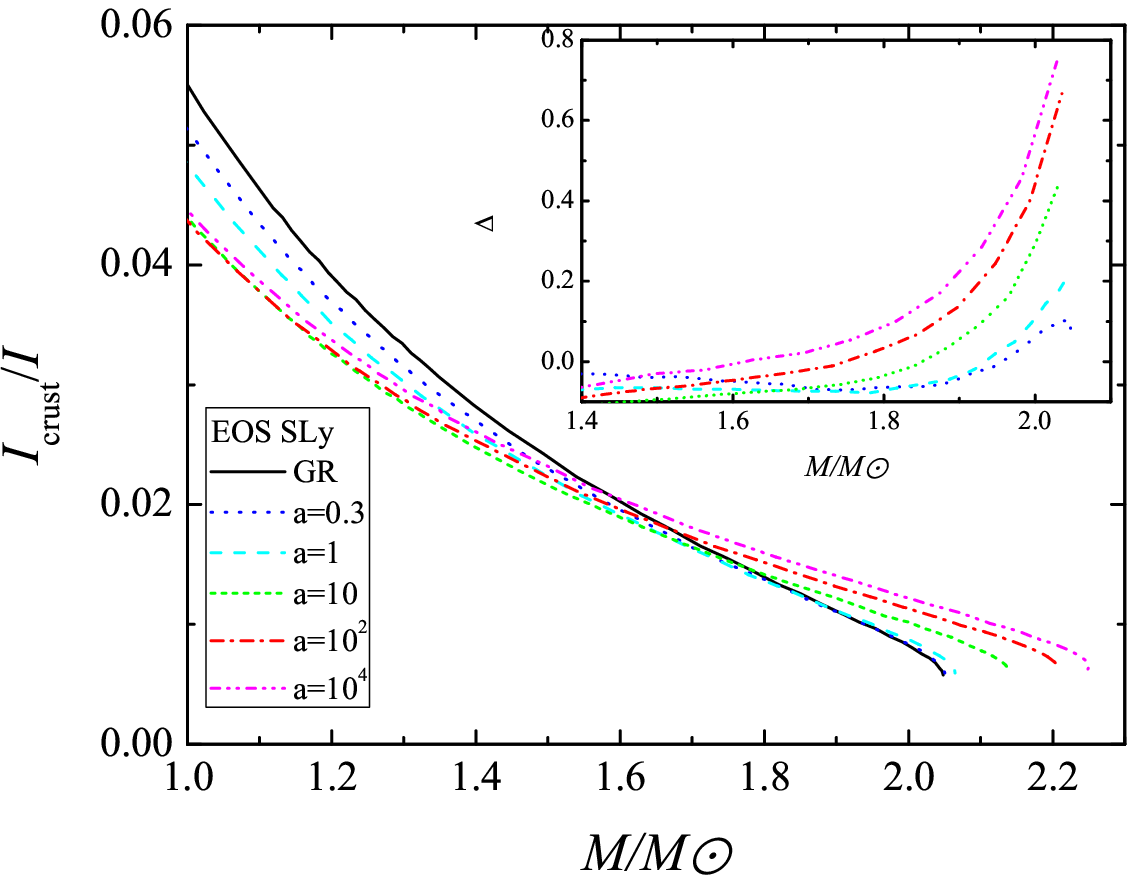}
\includegraphics[width=0.42\textwidth]{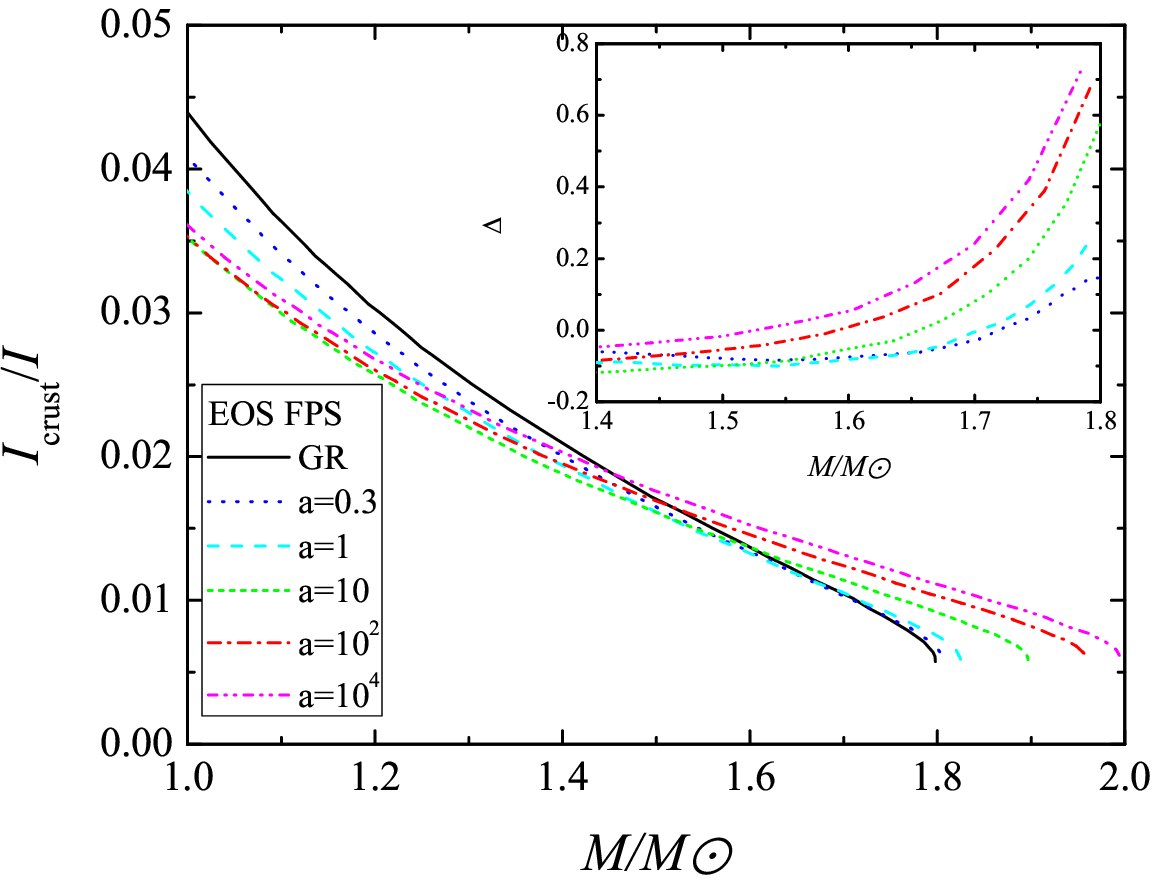}
\includegraphics[width=0.42\textwidth]{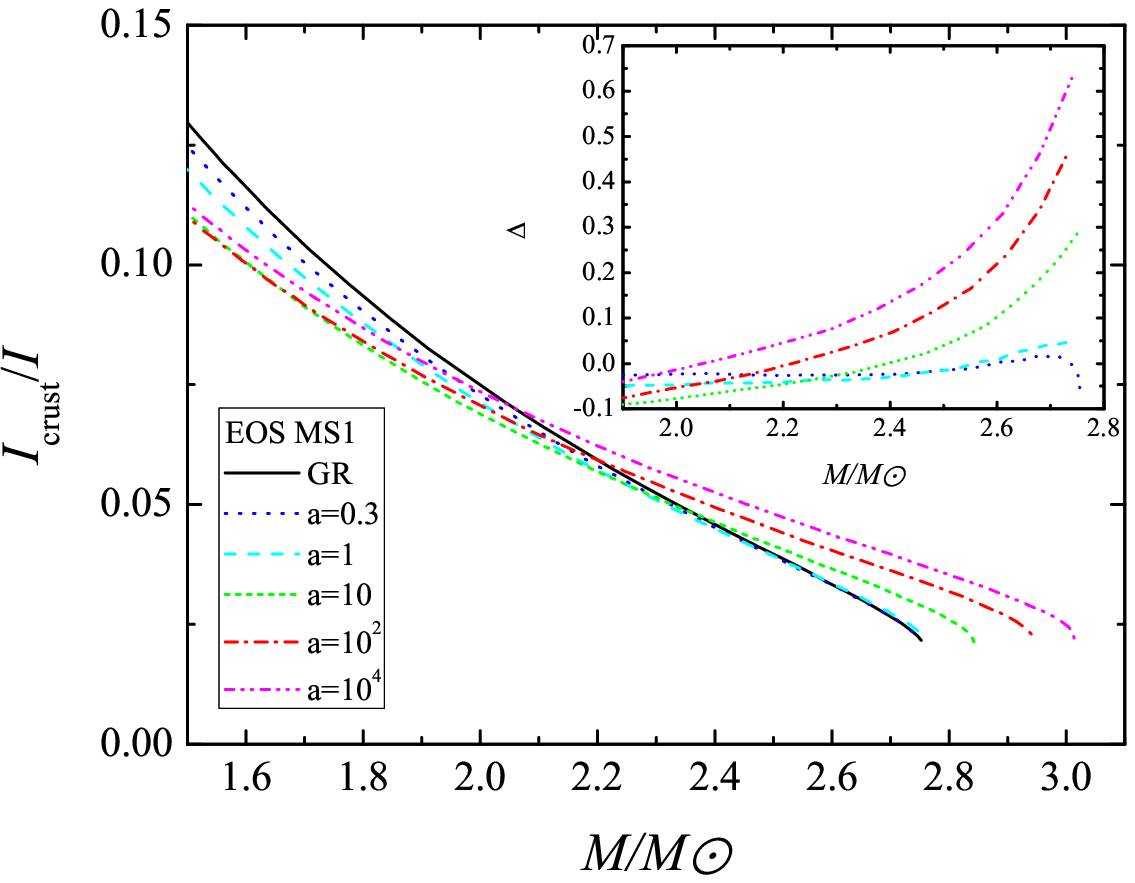}
\caption{Crust to the total moment of inertia ratio as a function of the mass of the neutron star in the case of ${\cal R}+a {\cal R}^2$ gravity. GR solutions are in continuous black line and  different colours and patterns correspond to different values of $a$. The inset shows the discrepancy between pure GR and the results in ${\cal R}^2$ gravity for different values of the parameter $a$.}
\label{Fig_crust_to_core}
\end{figure*}

We find that unlike STT the ${\cal R}^2$ gravity leads to serious deviations from pure general relativity as presented in Fig. \ref{Fig_crust_to_core} for a range of values of $a$.
For all EoS', we observe qualitatively the same behaviour: for low (high) masses the ratio is the highest (lowest) for GR, the cross-over occurring at around $M \simeq 1.5 M_\odot$.

In the insets we present the discrepancy between pure GR and the results in ${\cal R}^2$ gravity for different values of the parameter $a$. The discrepancy we define  as
\begin{equation}
\Delta = \frac{\frac{I_{\rm crust}}{I}( f({\cal R})) - \frac{I_{\rm crust}}{I} ( GR)}{\frac{I_{\rm crust}}{I} (GR)},
\end{equation}
and the presented results are for models with large masses up to the maximal ones in GR for the corresponding EoS. The deviation from GR with the increase of $a$ is clearly visible. The insets show that the increase of the crustal to the total moment of inertia ratio in the ${\cal R}^2$ gravity for the maximal GR mass compared to the pure GR case is more than 50\%.

For ${\cal R}^2$ gravity the problem with the deficiency of the crustal moment of inertia is partially solved---for models with masses higher than, roughly, $1.5 M_{\odot}$, the ratio increases, with  $a$ compared to GR. For models with small masses, however, the ratio decreases with $a$, so the moment of inertia deficiency problem gets even worse.

\section{Conclusions}

Given the identified deficiency of fraction of crustal moment of inertia to the total moment of inertia of neutron stars ($I_{\rm crust}/I$)---a critical parameter in the models addressing the observed glitch phenomena---we have investigated this ratio in scalar tensor theory of gravity with the conformal factor
$A(\phi)=e^{\frac{1}{2}\beta\phi^2}$ and ${\cal R}$-squared gravity model ($f({\cal R})={\cal R}+aR^2$) to see whether the free parameters of these theories could account for the discrepancy.

In the case of STT  we have found that $I_{\rm crust}/I$ changes much less than 3 \% for $\beta=-4.5$, the highest value compatible with solar system tests. This is an expected result, if we take into account the fact that the differences for all neutron star parameters are relatively small.
This implies that the deficiency of  $I_{\rm crust}/I$ can not be accounted for by referring to STT.

In the case of ${\cal R}$-squared gravity model we have found that for masses $M \lesssim 1.5 M_{\odot}$  the ratio $I_{\rm crust}/I$ is even smaller than its value in GR with increasing values of $a$ which renders the crustal moment of inertia deficiency problem even more acute.
For larger masses, however, we observe that
$I_{\rm crust}/I$ is greater than its value in GR by some 20\% to 60\% depending on the value of the parameter $a$ and the mass. This result suggests that the crustal moment of inertia deficiency could be accounted only if all observed glitching pulsars have masses larger than $M \gtrsim 1.5 M_{\odot}$. It should, however, be emphasized that young glitching pulsars have measured masses $1.4~M_\odot$ at which the difference in GR and ${\cal R}$-squared gravity is the smallest.

We then conclude that  the two gravity models we have considered in this paper are not candidates for addressing the crustal deficiency problem. This does not exclude that the problem could be accounted for in different models of gravity. Nor we can exclude an astrophysical explanation of the problem such as the one given in Ref. \cite{Guegercinouglu2014}.   Additional studies of alternative theories should be done as to allow more general conclusions to be made. For example,  the  investigation of the problem  in the context of STT with a massive  scalar field is in progress and the results will be presented elsewhere.

\section*{Acknowledgements}

KYE and ASA acknowledge support from the scientific and technological council of Turkey (T{\" U}B{\. I}TAK)
with the project number 108T686, and MMT with project
number 112T105. This work was supported by the COST
Action MP1304. SY and KS would like to thank the Research Group Linkage Programme of the
Alexander von Humboldt Foundation for the support. The support by the Bulgarian NSF Grant DFNI T02/6, Sofia
University Research Fund under Grant 70/2015 is gratefully acknowledged. KYE thanks Erbil G\"ugercino{\u g}lu for his comments.


\bibliography{ms-crust}

\end{document}